\documentstyle[emulateapj]{article}

\newcommand{\be}{\begin{equation}}
\newcommand{\ee}{\end{equation}}
\newcommand{\ba}{\begin{array}}
\newcommand{\ea}{\end{array}}
\newcommand{\siml}{\lower4pt \hbox{$\buildrel < \over \sim$}}
\newcommand{\simg}{\lower4pt \hbox{$\buildrel > \over \sim$}}

\slugcomment{Accepted for publication in ApJ Letters}

\begin{document}

\title{On Radio Quiescence of Anomalous X-ray Pulsars \\
    and Soft Gamma-Ray Repeaters}

\author{Bing Zhang}
\affil{Department of Astronomy \& Astrophysics, Pennsylvania State
    University, University Park, PA 16803}

\begin{abstract}
In the hot environments of the anomalous X-ray pulsars and the soft
gamma-ray repeaters, as indicated by their luminous, pulsed, quiescent
X-ray emission, $\gamma$-rays generated from the inner gaps may have 
shorter attenuation lengths via two photon pair production than via
magnetic photon splitting. The AXP/SGR environments may not be
pairless, even if photon splitting could completely suppress one
photon pair production in super-strong magnetic fields, as conjectured
by Baring \& Harding. Two-photon pair production more likely occurs
near the threshold, which tends to generate low energy pairs that are
not energetic enough to power radio emission in the observed bands.
However, emission in longer wavelengths may not be prohibited in
principle if these objects are indeed magnetars. The so-called
``photon splitting deathlines'' are still valid for high magnetic
field pulsars which are much dimmer in X-rays, if $\gamma$-rays with
both polarization modes split. 
\end{abstract}

\keywords{gamma rays: theory - magnetic fields - pulsars: general
 - radiation mechanisms: general - stars: neutron}

\section{Introduction}

There is now growing evidence that the anomalous X-ray pulsars (AXPs)
and the soft gamma-ray repeaters (SGRs) are isolated neutron stars
which form a distinct group with respect to the conventional
rotation-powered radio pulsars. In the quiescent state,
AXPs/SGRs differ from normal pulsars by showing the
following peculiar characteristics: (1) long periods, $P$, and high
spindown rates, $\dot P$; (2) much more luminous pulsed X-ray
emission with $L_x \sim 10^{35}-10^{36}~{\rm erg~s^{-1}}$; (3) no
firmly detected radio emission. A popular model for these objects 
is the magnetar model (e.g. Duncan \& Thompson 1992), which
involves strongly-magnetized neutron stars with dipolar magnetic
field strengths at the surfaces of order $B_p \sim 10^{14}-10^{16}$.
The magnetic braking naturally interprets the timing data of AXPs/SGRs
(e.g. Kouveliotou et al. 1998). The luminous pulsed X-ray emission
may be explained as decay of super-strong magnetic fields
(Thompson \& Duncan 1996). Radio quiescence of magnetars has been
attributed to the possible function of an exotic QED process, magnetic
photon splitting, which may strongly suppress one-photon pair
production in magnetic fields above roughly the critical value,
$B_q= m_e^2c^3/e\hbar \simeq 4.4 \times 10^{13}$ G (Baring
\& Harding 1998; 2001). 
The relative orientation between the magnetic and the
rotational axes may be also essential to judge whether a delayed pair
formation front might form, and hence, whether a
high magnetic field pulsar (HBP) or a magnetar is radio quiet (Zhang
\& Harding 2000). 
Here we will show through a
rough estimation that, in the hot magnetar environments as inferred
from the observations, a previously un-explored effect, i.e., two
photon pair production, may not be completely suppressed by magnetic
photon splitting, even if the latter can completely suppress one
photon pair production. The AXP/SGR environments therefore may not be
necessarily pairless. We discuss the possible implications of such a
process in understanding the AXP/SGR/HBP phenomenology.

\section{One-photon pair production, magnetic photon splitting, and
two-photon pair production}

An isolated rotating neutron star is a unipolar generator, which
generates a large potential drop between the magnetic pole and the
polar cap edge, i.e., $\Phi \sim 
B_pR^3\Omega^2/2c^2 \sim 6.6\times 10^{12} ~{\rm V} B_{p,14} P_1^{-2}
R_6^3$, where $B_p$, $\Omega$, $P$ and $R$ are the polar cap
surface magnetic field strength, the rotational frequency, the period,
and the radius of the neutron star, respectively. Here the parameters
are normalized to a typical magnetar, and the convention
$Q=10^{n}Q_n$ has been used. The maximum achievable electron energy is
$\gamma_{p,m} \sim e\Phi/m_ec^2 \sim 1.3\times 10^7 B_{p,14}P_1^{-2}
R_6^3$. In reality, such a high energy may be not achievable
due to screening of the parallel electric fields by pairs, or
due to radiation self-reaction of the electrons. The typical primary
particle energy, $\gamma_p$, is model-dependent (e.g. Zhang, Harding
\& Muslimov 2000, for normal field pulsars).  

One-photon pair production (also called $\gamma-B$ process), i.e.,
$\gamma \rightarrow e^{+}e^{-}$, is a first order QED process which
only operates in the presence of a strong magnetic field. It is
commonly accepted to be the main source of pairs in pulsar
magnetospheres, and
the conditions for this process to cease (e.g. Zhang et al. 2000)
or to be suppressed (e.g. Baring \& Harding 1998; 2001)
conventionally define the so-called radio emission
``deathlines'' of pulsars. In the high magnetic field regime which
we are interested in, one photon pair production occurs as
soon as the threshold condition, $\epsilon \sin\theta=2$, is
fulfilled, where $\epsilon$ is the $\gamma$-ray energy 
in units of the electron's rest mass, $m_ec^2$, $\theta \sim l/\rho 
+ \vec{\theta}_0$ is the angle between the photon momentum and the
local magnetic field. Here we denote $l$ as the distance the photon
would travel 
if the initial injection angle is zero, $\rho$ as the field line
curvature, so that the angle $l/\rho$ defines 
the electron's moving direction at the emitting point, which we call
the propagation direction. The vector
$\vec\theta_0$ is the initial photon ejection pitch angle with respect
to the electron's moving direction which should be added to
the $l/\rho$ term according to the vector algebra. The angle
$\theta_0=| \vec \theta_0 |$ 
depends on the mechanism that generates the photon, e.g., $\theta_0
\sim \gamma_p^{-1}$ for inverse Compton, and $\theta_0\sim 0$ for
curvature radiation. When $\theta_0 \sim 0$, the impact angle $\theta$ is
solely determined by the propagation effect, and the one-photon 
pair production attenuation length can be approximated as 
\be
\lambda_{1p}(\epsilon)=2\rho \epsilon^{-1}\simeq 1.84\times 10^8 ~{\rm
cm}~ \epsilon^{-1} P^{1/2} r_{e,6}^{1/2}  \xi^{-1},
\label{lambda1p}
\ee
where a dipolar geometry, $\rho=9.2\times 10^7 ~{\rm cm}~
r_{e,6}^{1/2}P^{1/2}\xi^{-1}$, is assumed throughout the paper,
$r_e$ is the radius at which the $\gamma$-ray is emitted, and
$0<\xi=\Theta_s/\Theta_{\rm pc}<1$ is the 
ratio of the field line magnetic colatitude at the surface to the
polar cap angle. For $\theta_0 \sim \gamma_p^{-1}$, the in-situ
one-photon threshold energy is $\epsilon_{th,1p} \sim 2/\theta_0 \sim
2\gamma_p \sim 2\times 10^5 \gamma_{p,5}$, photons with energies above
which have a pair-production attenuation length $\lambda_{1p}(\epsilon 
>\epsilon_{th,1p})\sim 0$. Far below this in-situ threshold, $\epsilon 
\ll 2\times 10^5 \gamma_{p,5}$, photons usually propagate until
$l/\rho \gg \theta_0$ and the attenuation length is essentially
defined by (\ref{lambda1p}). For photons with $\epsilon \siml 2\times
10^5 \gamma_{p,5}$, the superposition between the two direction
vectors is important, and the $\theta_0$ effect can either decrease
{\em or increase} the attenuation length depending on whether the
photon ejection direction is on the convex or concave side of the
field line. 

Magnetic photon splitting, i.e., $\gamma\rightarrow\gamma\gamma$, is a
more exotic, third order QED 
process which only becomes prominent when the field strength is above
$B_q$. Under the weak linear vacuum 
dispersion limit, only one splitting mode, i.e., $\perp\rightarrow
\|\|$, can fulfill the energy and momentum conservation requirements
simultaneously (Adler 1971). However, Baring \& Harding (1998; 2001)
argued that in the special environment of a magnetar, strong vacuum
dispersion may arise, so that it is possible that all
the three photon splitting modes permitted by charge-parity invariance
in QED, i.e., $\perp\rightarrow \| \|$, $\perp\rightarrow \perp\perp$,
and $\|\rightarrow \perp\|$, may operate together. If this is indeed
the case, photons with both polarization modes split, and a splitting
cascade will occur in a magnetar inner magnetosphere until
all the hard photons split to soft ones below the one-photon
pair production threshold. The magnetar magnetosphere is therefore
essentially pairless, which may offer a natural explanation for radio
quiescence of the known AXPs/SGRs. In the low-energy non-dispersive
limit, the polarization-averaged photon splitting
attenuation coefficient is (Harding, Baring \& Gonthier 1997)
\begin{eqnarray}
T_{sp}(\epsilon) & \approx {\alpha^3\over 10\pi^2}{1\over \Lambda}
\left({19\over 315}\right)^2  {\cal F}(B') {B'}^6 \epsilon^5
\sin^6 \theta 
\nonumber \\
& \simeq 0.37 {\cal F}(B') {B'}^6 \epsilon^5
\sin^6 \theta,
\label{Tsp}
\end{eqnarray}
where $\alpha$ is the fine-structure constant, $\Lambda=\hbar /(m_ec)
\simeq 3.86\times 10^{-11}$ cm is the Compton wavelength of the
electron, $B'=B/B_q$, and ${\cal F}(B')$ is a 
strong-field modification factor, which $=1$ for $B'<<1$, and $\simeq
(1/12)\times (315/19)^2\times 3 \times (1/6B'^3)^2 \simeq 1.9
{B'}^{-6}$ for $B'>>1$ (Harding et al. 1997). Generally, we can assume
${\cal F}(B') \simeq {\cal A} {B'}^{-\beta}$ in a certain $B'$ regime,
where ${\cal A}$ and $\beta$ are constants in that regime. Again, for
$\theta_0=0$, one can solve for the averaged photon splitting
attenuation length, $\lambda_{sp}$, through the definition
$\int_0^{\lambda_{sp}(\epsilon)} T_{sp}(\epsilon,s)  
ds=1$, where $s$ is the trajectory length of the photon propagation.  
Neglecting $B'$ declination with 
height which is not important for $\epsilon \gg 1$, this gives
$\lambda_{sp}(\epsilon)\simeq 1.5 {\cal A}^{-1/7} {B'}^{-(6-\beta)/7} 
\epsilon^{-5/7} \rho^{6/7}$. For a pure dipolar geometry, this is
\be
\lambda_{sp}^{B'\sim 1}(\epsilon) \simeq 
1.0\times 10^7 ~{\rm cm}~ \epsilon^{-5/7} {B'}^{-15/28} P^{3/7}
r_{e,6}^{3/7} \xi^{-6/7}
\label{lambdasp1}
\ee
for $B'\simg 1$ (relevant to HBPs), where
$\beta\simeq 9/4$ has been adopted based on an approximate fit to the
numerical calculations of 
Baring \& Harding (2001). When $B' \gg 1$ which is more relevant to
AXPs/SGRs, ${\cal A}=1.9$, $\beta=6$, and the $B'$-dependence
disappears, which gives 
\be
\lambda_{sp}^{B'\gg 1}(\epsilon)\simeq 0.9\times 10^7
~{\rm cm}~ \epsilon^{-5/7} P^{3/7}r_{e,6}^{3/7} \xi^{-6/7}. 
\label{lambdasp2}
\ee
For $\theta_0 \neq 0$, the above 
approximations are still valid far below the in-situ one-photon pair
production threshold.

Two-photon pair production, i.e., $\gamma\gamma\rightarrow
e^{+}e^{-}$, is a second-order process and can execute even without
the presence of a magnetic field. Its relative importance with
respect to the one photon pair production within the
context of a neutron star has been studied previously (e.g. Burns \& 
Harding 1984; Zhang \& Qiao 1998). In the polar cap context,
the interaction is between the two photon species, i.e.,
a soft X-ray component due to the thermal emission from the surface,
and a hard $\gamma$-ray component produced by the relativistic primary
electrons via curvature radiation or inverse Compton scattering.
The importance of this process generally requires a hot polar cap
(Zhang \& Qiao 1998), which seems not achievable in normal pulsars
(e.g. Harding \& Muslimov 2001). However, AXPs/SGRs
have luminous quiescent X-ray emission, and their spectra usually
include a hot blackbody component (e.g. $kT \sim 0.43$ keV$\sim
5\times 10^6$ K for 1E 2259+586) with a much larger inferred emission
area (the radius $\sim 3$ km for 1E 2259+586, e.g. Corbet et
al. 1995; Rho \& Petre 1997). Both the high temperature and the large
emission area make two-photon pair production a potentially important
process to explore.
The threshold condition for magnetic two-photon pair production is
(Daugherty \& Bussard 1980; Kozlenkov \& Mitrofanov 1986)
\be
(\epsilon_1 \sin\theta_1 +\epsilon_2 \sin\theta_2)^2 +
2\epsilon_1\epsilon_2 [1-\cos(\theta_1-\theta_2)] \geq 4,
\label{2p}
\ee
where $\epsilon_1$ and $\epsilon_2$ are the soft and hard photon
energies, and $\theta_1$ and $\theta_2$ are the angles of their
directions of momenta with respect to the magnetic field line,
respectively. In the present problem, we have $\epsilon_2 \gg
\epsilon_1$ ($\epsilon_2/\epsilon_1 \sim 10^6$ at the threshold).
Given $\theta_2 = \gamma_p^{-1}$,  the first term in (\ref{2p}) is
negligible as long as $\gamma_p \gg 10^3$, and the threshold condition
is reduced to that of 
non-magnetic two-photon pair production, i.e., $\epsilon_1\epsilon_2
(1-\cos\theta_1) \geq 2$ (e.g. Gould \& Schr\'eder 1967).
For a semi-isotropic blackbody emission from surface, we further have
$\theta_1 \siml 90^{\rm o}$. The typical
soft photon energy is $\epsilon_1 \simeq \Theta = kT /m_ec^2 = 8.4
\times 10^{-4} (T_6/5)$. Hard photons with $\epsilon_2 \geq
\epsilon_{th,2p}=\Theta^{-1}\simeq 1.2 \times 10^3 (T_6/5)^{-1}$ can in
principle pair produce in the soft photon sea. These photons are
readily generated by primary particles via inverse Compton off the
thermal photons and the photon splitting cascade (\S3).

What is more essential is to estimate the $\gamma$-ray attenuation
length of this process. Though a fully magnetic 
two-photon pair production treatment (e.g. Kozlenkov \& Mitrofanov 
1986) is desireble, no directly usable 
formula for the present problem (two photon species) is available at 
this time. We therefore treat the process using the non-magnetic 
cross section, keeping in mind that the strong field effect may
influence some of the conclusions considerably.
The non-magnetic polarization-averaged $\gamma$-ray attenuation 
coefficient with energy $\epsilon$ in a thermal photon sea 
characterized by $\Theta = kT/m_ec^2$ can be written as (Gould 
\& Schr\'eder 1967; Zhang \& Qiao 1998) 
\be
T_{2p}(\epsilon,s)=\frac{\alpha^2}{\pi \Lambda} \Theta^3 F(\epsilon,s)
\simeq 2.1\times 10^{-6} T_6^3 F(\epsilon,s),
\label{T2p}
\ee
where $s$ is the trajectory length of the photon assuming the photon
being ejected at the surface. We may write the 
reduction factor $F(\epsilon,s)=g(s)f(\epsilon)$, where $g(s)\simeq
(0.270 -0.507\mu_c+0.237\mu_c^2)$ takes care of the non-isotropy of
the soft photons with respect to the height (Zhang \& Qiao 
1998), and $\mu_c=\cos\theta_c=[1-R^2/(R+s)^2]^{1/2}$ is the
maximum cosine of the impact angle between the two photons, which is
applicable when the hot surface area is much larger than the polar
cap as is the case for AXPs/SGRs. For $s \ll R$, and noticing the
semi-isotropic surface emission, we have $\mu_c \sim 0$ and $g(s) \sim
0.27$. The function $f(\epsilon)$ reaches  
the maximum  $\sim 1$ when $\epsilon \sim \epsilon_{th,2p}\sim
\Theta^{-1}$, and declines as $\epsilon^{-1}$ when $\epsilon \gg
\epsilon_{th,2p}$ in the form of $f(\epsilon) \sim (\pi^2/3)\Theta^{-1}
\epsilon^{-1} \ln (0.117 \Theta\epsilon)$ (Gould \& Schr\'eder 1967). 
Near $\epsilon_{th,2p}$, the attentuation length for two
photon production is then
\be
\lambda_{2p}(\epsilon_{th,2p}) \sim [T_{2p}(\epsilon_{th,2p})]^{-1}
\simeq 1.4\times 10^4 ~{\rm cm}~ (T_6/5)^{-3}.
\label{lambda2p}
\ee
At the same $\gamma$-ray energy, i.e., $\epsilon_{th,2p} \sim 1.2\times
10^3$ ($T_6\sim 5$ adopted), we have $\lambda_{sp}(\epsilon_{th,2p}) \sim 
1.5\times 10^5~ {\rm cm}~ P_1^{3/7}$
(eq.[\ref{lambdasp2}]), and $\lambda_{1p} (\epsilon_{th,2p}) \sim 4.9
\times 10^5~ {\rm cm}~ P_1^{1/2}$
(eq.[\ref{lambda1p}]). We therefore have
\be
\lambda_{2p}(\epsilon_{th,2p}) < \lambda_{sp}(\epsilon_{th,2p}) 
< \lambda_{1p}(\epsilon_{th,2p})
\label{2p<sp<1p}
\ee
for a typical AXP/SGR environment.
This means that even if photon splitting can suppress one-photon pair 
production, it cannot suppress two-photon pair production, at least
at the two-photon threshold. Several comments ought to be made. 
First, again strong fields may modify the conclusion greatly. Though
more detailed investigations are required to tell these modifications, 
we may try to guess some of these effects. Generally, strong fields
will suppress the two-photon cross section, but there are
``resonances'' at which the cross sections are much enhanced, due
to discreted Landau energy levels of the created pairs (Kozlenkov \&
Mitrofanov 1986). Given the broad spectral distributions of both the
hard photons and the soft photons, there may always be some preferred
two-photon attenuations to occur with even shorter attenuation lengths
than the one estimated in (\ref{lambda2p}), so that (\ref{2p<sp<1p})
may be not changed. Second, equation (\ref{2p<sp<1p}) is
applicable when $\theta_0=0$, or, when $\epsilon \ll \epsilon_{th,1p}$ 
if $\theta_0\neq 0$.
Third, in all the above treatments, polarization-averaged cross
sections have been adopted, while the threshold conditions for both 
one-photon and two-photon pair production are polarization-dependent. 
Nonetheless, usually the primary $\gamma$-rays are dominated by the
$\perp$ mode, and polarization-dependent treatments will not change the 
above discussions qualitatively.
Finally, two photon pair production attenuation length is sensitive to 
the surface temperature ($\propto T^{-3}$). The condition to ignore the 
two-photon process is $\lambda_{2p}(\epsilon_{th,2p}) > {\rm max}
[\lambda_{sp}(\epsilon_{th,2p}), \lambda_{1p}(\epsilon_{th,2p})]$, or
\be
T_6 < {\rm max} ~(3.4 P^{-3/26}, 2.7 P^{-1/8}). 
\label{T6}
\ee
A non-zero $\theta_0$, or a smaller hot area (e.g. polar cap in normal 
pulsars) will make the constraint less stringent. The condition
(\ref{T6}) usually holds for HBPs, but not for AXPs/SGRs.

\section{Radio quiescence of AXPs/SGRs and deathline for HBPs} 

Pair production in a pulsar magnetosphere is believed to be an
essential, though perhaps not sufficient, condition for coherent
radio emission. Radio quiescence of the AXPs/SGRs has been attributed
by Baring \& Harding (1998; 2001) to the suppression of pair
production by photon splitting, under the hypothesis that photons with both
polarization modes split. However, as discussed above, two-photon
process is another pair formation mechanism, and it may still produce
pairs even if the photon splitting hypothesis holds, at least
for photons near the two-photon threshold with a zero or negligible
impact angle with the field line. The question is whether a copious
number of such photons exist in the magnetar inner magnetosphere. 

The typical energy of the primary particles in a magnetar environment
is not well-studied. Nonetheless, copious pairs may be not generated
initially, so it is reasonable to expect $\gamma_p$ not too much
below $\gamma_{p,m}\sim 10^7$. Since magnetars are slow rotators, 
characteristic curvature radiation energy,
$\epsilon_{cr}=(3/2)(\gamma_p^3 \hbar c/
\rho m_ec^2) \sim 200 P_1^{-1/2} \gamma_{p,7}^3$, is below
$\epsilon_{th,2p}$, even for the maximum electron energy. 
The main source of the hard photons above 1 GeV is inverse Compton
scattering (Gonthier et al. 2000).
Resonant scattering off the thermal peak generally requires
$\gamma_p \sim B'/\Theta \sim 6\times 10^4 B'_1$, a condition easily
satisfied during the acceleration phase of the primary particles. The
characteristic photon energy is $\epsilon_{res} \sim \gamma_p B' \sim
6\times 10^5 (B'_1)^2$, which is well above $\epsilon_{th,2p}$ and
around $\epsilon_{th,1p}$. These $\gamma$-rays may otherwise pair
produce via one-photon process, but under the Baring \& Harding photon
splitting hypothesis, they 
will split to lower energy photons through a splitting cascade. 
The daughter photon energies approach 
$\epsilon_{th,2p}$ after several generations where the two-photon
attenuation length is the minimum. Assuming that the daughter photons
mainly follows the direction of 
the parent photon, The $\theta_0$ effect at this time
is not important since $\epsilon_{th,2p} \ll \epsilon_{th,1p}$, and
according to (\ref{2p<sp<1p}), we expect that at least some pairs will
produce via the two-photon process. If these pairs are not copious
enough to screen the parallel electric field, electrons keep
accelerating to reach higher energies, e.g., $\gamma_p \sim
10^6$. These electrons will also inverse Compton scatter the thermal
photons in the Klein-Nishina regime with typical photon energy
$\epsilon_{_{KN}} \sim 10^6 
\gamma_{p,6}$ and the initial ejection angle $\gamma_0 \sim 10^{-6}
\gamma_{p.6}^{-1}$. These photons may also undergo photon splitting
cascade, and pair produce near $\epsilon_{th,2p}$. This goes on 
until primary electron energy saturates.

At present, there is no firm detection of pulsed radio emission from
the known AXPs and SGRs. 
This is a natural expectation if these objects are accretors
(e.g. Chaterjee, Hernquist \& Narayan 2000). However, if they are
indeed magnetars, the non-detection of emission in the
``conventional'' radio band ought be re-investigated due to the above
reason. It could be possible that the hot environments may destroy some
fragile conditions (e.g., bunching condition or maser condition) that
are necessary to generate coherent radio emission. Other suggestions
include the influence of the bursting activities (Thompson 2000). 
Here we propose that even if
photon splitting can completely suppress one-photon pair production,
radio emission from AXPs/SGRs may be not prohibited. The
emission, if at all, should be mainly emitted in longer wavebands. For
example, the characteristic curvature radiation frequency of the pairs
is $\nu_{cr} 
=(3/4\pi) (c/\rho) \gamma_\pm ^3 \propto P^{-1/2} \gamma_\pm^3$, where
$\gamma_\pm$ is the Lorentz factor of the pairs. In AXPs/SGRs, both
long periods and small $\gamma_\pm$'s tend to lower the typical radio
emission frequency. The latter is mainly due to that pairs are more
likely to generate near $\epsilon_{th,2p}$. We thus expect that the
typical pair energy in AXPs/SGRs might be less than that in normal 
pulsars generated via one-photon pair production, and that
AXPs/SGRs may be radio loud in lower frequency bands. There are
reports that SGR 1900+14 (Shitov 1999) and the AXP 1E 2259+586
(Malofeev \& Malov 2001) have been detected with low-frequency pulsed
emission at 111 MHz. If these 
claims are confirmed, they are consistent with the theoretical
expectation presented here. Also Gaensler et al. (2001) recently
pointed out that non-detection of radio emission in AXPs/SGRs does
not necessarily mean that they are intrinsically radio quiet due to
the inadequate searching sensitivity and the beaming effect.

For HBPs, (\ref{T6}) generally holds, and one can safely neglect 
two-photon pair production. The
photon-splitting-dominant condition is then $\lambda_{sp}(\epsilon)
<\lambda_{1p}(\epsilon)$, which translates to (with the use of
eqs.[\ref{lambda1p}] and [\ref{lambdasp1}]) 
${B'}^{15/28} \epsilon^{-2/7} P^{1/14} r_{e,6}^{1/14} \xi^{-1/7} >
0.054$ with a negative $\epsilon$-dependence.
Baring \& Harding (1998; 2001) defined a
``photon splitting deathline'' using the criterion that the escape
energy of both one-photon pair production and photon splitting 
is equal. More generally, one may derive the typical photon
energy $\epsilon$ from a gap as well as its $P-$, $B_p-$dependences by 
choosing a gap boundary condition and a $\gamma$-ray emission
mechanism (e.g. Zhang et al. 2000) to define the ``photon splitting
dominant line'' for a particular model. These lines 
generally tilt up in the short period regime relative to the Baring
\& Harding (1998) deathline, and allow some short-period HBPs, e.g.,
PSR J1119-6127 (Camilo et al. 2000), to be below the line. This is
understandable: Faster pulsars tend to have larger acceleration
potentials and hence, produce more energetic photons than the slower
pulsars, and for higher energy photons, even higher magnetic fields
are required for photon splitting to dominate. For example, for a
resonant inverse 
Compton-controlled vacuum gap (which is more possible in the high B
regime), the typical photon energy may be expressed as
$\epsilon_c({\rm ICS-VG}) \simeq 5.5\times 10^4 P^{1/14} {B'}\xi^{-3/7}$,
and the photon splitting dominant line is (Zhang \& Harding 2001)
\be
B_p \geq 1.0\times 10^{14} {\rm G} (P/ 1 {\rm s})
^{-10/49}\xi^{4/49}, 
\label{1}
\ee
or $\dot P\geq 2.44\times 10^{-12} (P/ 1 {\rm s}) ^{-69/49}
\xi^{8/49}$ by adopting $B_p=6.4\times 10^{19} {\rm G} \sqrt{P\dot
P}$. This is also the deathline for 
the anti-parallel rotators in which vacuum-type inner gaps are
expected. For parallel rotators, a similar line for the
space-charge-limited-flow type inner gap may be obtained after detailed
numerical calculations (Harding et al. 2001, in
preparation). This is a line to define whether delayed pair formation
is necessary, rather than a deathline, which is then defined according 
to the binding condition at the surface (Zhang \& Harding 2000).

\section{Summary \& discussion}
We have shown that if AXPs/SGRs are indeed magnetars, even if photon
splitting could completely suppress one-photon pair production, 
$\gamma$-rays in the magnetar magnetosphere may still generate
electron-positron pairs via two-photon pair production,
mainly because the AXP/SGR environments are very hot. Non-detection of 
radio emission from AXPs/SGRs may be because the low energy pairs
generate radio emission with too low a frequency to be observable in the 
bands above several hundred MHz. Searching for
low frequency emission from these objects is of great interest, and if 
detected, the low frequency emission will rule out the accretion
models. This conclusion does not expel the 
photon splitting hypothesis which could be tested in HBPs. For example,
if the recently discovered 424 ms radio quiet pulsar (Zavlin
et al. 2000) turns out a HBP, it may lend support to the photon
splitting hypothesis of Baring \& Harding (1998), and the geometric
proposal of Zhang \& Harding (2000). 

Several caveats ought to be noted.
Only a very crude treatment, especially for the two-photon pair
production, is performed in this letter. More detailed cascade 
simulations by including all the three relevant QED processes, i.e.,
one-photon, two-photon pair production, and photon splitting, with a
full ``magnetic'' treatment are needed to provide a clearer understanding
of the whole phenomenon. As a first step, a more user-friendly
treatment of the two-species magnetic two-photon pair production is
called for. Dipolar magnetic fields have been assumed throughout.
In magnetars, multi-pole components may exist, and this will enhance
one-photon production and may weaken the proposal discussed
here. Finally, two photon pair production may also be important in
another type of pulsars, i.e., the millisecond pulsars, where the one
photon process is less important due to the weak dipolar fields
involved. 

\acknowledgements
I am grateful to Alice Harding, Matthew Baring and the anomalous
referee for insightful comments, to Peter M\'esz\'aros, George
Pavlov and RenXin Xu for discussions and encouragements, and to NASA
(NAG5-9192 and NAG5-9153) for supports.

\end{document}